# Theory of angle-resolved photoemission experiments on a two-band model

Tian De Cao[*]

*Department of Physics, Nanjing University of Information Science & Technology, Nanjing 210044, China*

**Abstract**

Considering the electron states inside and outside the solid, we derive a formula of photoemission intensity. A general theoretical way to determine electronic structures of solids from ARPES experiments is outlined. It is shown that the spectral function inside the solids cannot be measured directly by ARPES, the effects of free electron states on the electronic structure observed by ARPES measurements must be considered, and the results from ARPES experiments cannot be understood until these results have been made consistent with a theoretical calculation.

## 1. Introduction

Angle-resolved photoemission spectroscopy (ARPES) is accepted as a leading tool in the investigation of high-Tc superconductors [1], and a lot of works about ARPES using a one-band model have been reported [2-5]. We have observed that ARPES experiments contain very complex phenomena, some of which can be attributed to the ones known in other experiments. For example, a photoelectron associated with the deeper core-levels can be discriminated from the one associated with the Bloch bands, and an electron state inside the material must be distinguished from the one on the surface. Although the electron states inside the solids could be observed with photoelectrons, there is no method to clarify how to eliminate the effects of the electron states outside the solids.

In fact, the usual suggestion in literatures is that the electron spectral function within the solid is directly observed by ARPES, but we find that this suggestion is not appropriate for the spectral functions of electron systems in strongly correlated materials. This work derives the formula of photoemission intensity on a two-band model. It is shown that the electronic structures in the strongly correlated materials must be determined on the basis of combining the ARPES experiment with the solution of the theoretic model like the periodic Anderson model. This means that ARPES experiments cannot be understood until they have been made consistent with a theoretic calculation.

---

[*]Corresponding author.
[*]E-mail address: tdcao@nuist.edu.cn (T. D. Cao).
[*]Tel: 011+86-13851628895

2. **Formula**

In order to find the electronic structure below the Fermi energy with the photoelectrons outside the solid, the wave functions both within and outside the solid must be described with the same Hamiltonian. Provided the effects of the surface of a crystal could be eliminated from the results observed in experiments, we consider the electron states both inside and outside the solid, and take the Hamiltonian form

$$H = H_0 + H_{int} \tag{1}$$

The interaction with photon is treated as the perturbation given by

$$H_{int.} \approx \frac{je\hbar}{m}\vec{A}\cdot\vec{\nabla} \tag{2}$$

where $\vec{\nabla}\cdot\vec{A}=0$, $\vec{A}$ is the external classical vector potential. A common basis set for the starting point of the many-body calculation is just free-particle wave function $\psi(\vec{x}) = \sum_{\vec{k}} c_{\vec{k}} e^{j\vec{k}\cdot\vec{x}}$ for good metals. However, the wave functions both within and outside the strongly correlated material should be described with other basic functions on the basis of the so-called superposition principle of states. These wave functions could be

$$\psi(x) \equiv \psi(\vec{x},s) = \sum_{\vec{q},\sigma} d_{\vec{q}\sigma}\phi_{\vec{q}}(\vec{x})\chi_\sigma + \sum_{\vec{k},\sigma} c_{\vec{k}\sigma} e^{j\vec{k}\cdot\vec{x}}\chi_\sigma \tag{3}$$

where $d_{\vec{q}\sigma}$ destroy an electron in $\vec{q}$ state of spin $\sigma$ inside the solid, and $c_{\vec{k}\sigma}$ destroy an electron outside the solid. Thus we can take the form

$$H_0 = \sum_{\vec{k},\sigma} E_k c^+_{\vec{k}\sigma} c_{\vec{k}\sigma} + \sum_{\vec{k},\sigma}(M_k d^+_{\vec{k}\sigma} c_{\vec{k}\sigma} + h.c.) + \sum_{\vec{k},\sigma} \varepsilon_k d^+_{\vec{k}\sigma} d_{\vec{k}\sigma} + H_{d-d} \tag{4}$$

where $H_{d-d}$ represent the interactions between electrons within the material, while the multi-band model, such as the one for cuprate superconductors, requires

$$\psi(x) \equiv \psi(\vec{x},s) = \sum_{\vec{q},\sigma} d_{\vec{q}\sigma} \phi_{\vec{q}}(\vec{x}) \chi_\sigma + \sum_{\vec{q},\sigma} p_{\vec{q}\sigma} \varphi_{\vec{q}}(\vec{x}) \chi_\sigma + \sum_{\vec{k},\sigma} c_{\vec{k}\sigma} e^{j\vec{k}\cdot\vec{x}} \chi_\sigma \qquad (5)$$

and so on. This will give more complex model than Eq.(4). However, a 'multi-band' model could be reduced into a one-band model for the electronic properties of these materials, thus our discussion is limited to the model (4). Moreover, Eq.(6)-(15) below can be extended to other models.

We can apply the Golden Rules in quantum mechanics to the many-particle theory, and write the transition rate between the N-electron initial state $|\psi_i^N>$ and the final state $|\psi_f^N>$ in

$$w_{fi} = \frac{2\pi}{\hbar} |<\psi_f^N|H_{int.}|\psi_i^N>|^2 \delta(E_f^N - E_i^N - h\nu) \qquad (6)$$

It should be noted that both the final state $\psi_f^N$ and the initial state $\psi_i^N$ are the states associated with the same Hamiltonian $H_0$, thus $\psi_f^N$ is one of all possible initial states $\{\psi_i^N\}$ in Eq.(6). Because the effects of surface states are eliminated, the wave vector $\vec{k}$ is seen as the conserved one. Particularly, this should be reasonable when $\vec{k}$ is approximately parallel to the surface of the solid.

We write

$$<\psi_f^N|H_{int}|\psi_i^N> = <\phi_f^{\vec{k}}|H_{int}|\phi_i^{\vec{k}}><\psi_m^{N-1}|\psi_i^{N-1}> \qquad (7)$$

as discussed in the literatures, the final state $\phi_f^{\vec{k}}$ and the initial state $\phi_i^{\vec{k}}$ for single particle can be signed by the wave vector $\vec{k}$, thus we rewrite Eq.(7) as

$$<\psi_f^N|H_{int}|\psi_i^N> = M_{int}^{\vec{k}} <\psi_m^{N-1}|\psi_i^{N-1}> \qquad (8)$$

where $f$ expresses $(\vec{k},m)$. Because $\sum_k w_{fi}|_{alli} = V\int \frac{d^3k}{(2\pi)^3} w_{fi}|_{alli}$ and $d^3k = k^2 dk d\Omega = \frac{m_e}{\hbar^2} k dE_{kin} d\Omega$ for the free electron states outside the solid, the counting rate of the photoelectrons per second per solid angle per energy interval is expressed in the form

$$I(\vec{k},E_{kin}) = Ck\sum_{i,f} p_i |<\psi_f^N|H_{int.}|\psi_i^N>|^2 \delta(E_f^N - E_i^N - h\nu)$$

$$= |M_{int}^{\vec{k}}|^2 k \sum_{i,m} p_i |<\psi_m^{N-1}|\psi_i^{N-1}>|^2 \delta(E_f^N - E_i^N - h\nu) \qquad (9)$$

where $p_i = e^{-\beta(E_i^N - \mu N - \Omega)}$ is the probability of one initial state, and $\sum_{i,m}$ do not contain the summation over the wave vectors $\vec{k}$. It is necessary to note that $<\psi_m^{N-1}|\psi_i^{N-1}> \neq \delta_{m,i}$ because $|\psi_i^{N-1}>$ and $|\psi_f^{N-1}>$ could not be the eigenfunctions of the same Hamiltonian. Since each transition from the initial state to the final one has a photoelectron which has wave vector $\vec{k}$, thus we can write $|\psi_i^{N-1}> \sim a_{\vec{k}\sigma}|\psi_i^N>$, $a_{\vec{k}\sigma}$ should be neither $d_{\vec{k}\sigma}$ nor $c_{\vec{k}\sigma}$ because $|\psi_i^N>$ is not the eigenfunction of $d_{\vec{k}\sigma}^+ d_{\vec{k}\sigma}$ or $c_{\vec{k}\sigma}^+ c_{\vec{k}\sigma}$. The $a_{\vec{k}\sigma}$ is the quasi-particle destruction operator, and $|\psi_i^N>$ is the eigenfunction of $a_{\vec{k}\sigma}^+ a_{\vec{k}\sigma}$. Eq.(8) becomes

$$I(\vec{k}, E_{kin}) = C|M_{int}^{\vec{k}}|^2 k \sum_{m,i} p_i |<\psi_m^{N-1}|a_{k\sigma}|\psi_i^N>|^2 \cdot \delta(E_m^{N-1} + E_{kin} + \phi - E_i^N - h\nu) \tag{10}$$

where $E_m^{N-1}$ corresponds to the energy of the (N -1)-electron state $\psi_m^{N-1}$, other symbols follow Ref. [1]. The expression form of the intensity can be changed from Schrödinger to Heisenberg time-dependent operators with the aid of the integral

$$\delta(E_f^N - E_i^N - h\nu) = \frac{1}{2\pi\hbar} \int_{-\infty}^{+\infty} d(t-t') e^{j(E_f^N - E_i^N - h\nu)(t-t')/\hbar} \tag{11}$$

and this implies

$$|<\psi_m^{N-1}|a_{\vec{k}\sigma}|\psi_i^N>|^2 \delta(E_m^{N-1} - E_i^N + E_{kin} + \phi - h\nu)$$
$$= \frac{1}{2\pi\hbar} \int_{-\infty}^{+\infty} d(t-t') e^{j(E_{kin}+\phi-h\nu)(t-t')/\hbar} <\psi_m^{N-1}|a_{\vec{k}\sigma}(t)|\psi_i^N> \times <\psi_i^N|a_{\vec{k}\sigma}^+(t')|\psi_m^{N-1}> \tag{12}$$

Using Eq. (12), we rewrite Eq.(10) as

$$I(\vec{k}, E_{kin}) = Ck \frac{1}{2\pi\hbar} \int_{-\infty}^{+\infty} d(t-t') e^{j(E_{kin}+\phi-h\nu)(t-t')} \times \sum_i e^{-\beta(E_i^N - \mu N - \Omega)} <\psi_i^N|a_{\vec{k}\sigma}^+(t) a_{\vec{k}\sigma}(t')|\psi_i^N> \tag{13}$$

Introducing the function

$$G_a^<(\vec{k},\sigma,t-t') = \sum_i e^{-\beta(E_i^N - \mu N - \Omega)} <\psi_i^N|a_{\vec{k}\sigma}^+(t) a_{\vec{k}\sigma}(t')|\psi_i^N> \tag{14}$$

one will find that this function is related to the spectral function [6] by

$$G_a^<(\vec{k},\sigma,\omega) = j n_F^a(\omega) A_a(\vec{k},\sigma,\omega)$$

thus we get the expression

$$I(\vec{k}, E_{kin}) = C' M_{int}^2 k\, n_F^a(E_{kin} + \phi - h\nu) \cdot A_a(\vec{k}, \sigma, E_{kin} + \phi - h\nu) = C' M_{int}^2 k\, n_F^a(E_B^{\vec{k}})\, A_a(\vec{k}, \sigma, E_B^{\vec{k}}) \quad (15)$$

where $E_B^{\vec{k}}$ is the energy of electron ($E_B^{\vec{k}} = 0$ at $E_F$) inside the solid. In the expressions above, we have used the total energy conservation $E_{kin} + \phi = h\nu + E_B^{\vec{k}}$ as given in Ref. [1]. It seems that Eq. (15) is similar to the form in the literatures, but it is shown that what ARPES experiments measure is the spectral function $A_a$ of quasi-particles which are related to the electrons both inside and outside the solid, instead of the one suggested by some authors.

## 3. Results and discussions

However, we will find that the electronic structures of the solid can be understood in Eq. (15). To find the spectral function $A_a(\vec{k}, \sigma, \omega)$, we should calculate the Green's function

$$G_a(\vec{k}, \sigma, \tau - \tau') = -<T_\tau a_{\vec{k}\sigma}(\tau) a_{\vec{k}\sigma}^+(\tau')> \quad (16)$$

with which we get the retarded function

$$G_{a,ret}(\vec{k}, \sigma, \omega) = \underset{j\omega_n \to \omega + i\delta}{change} G_a(\vec{k}, \sigma, j\omega_n) \quad (17)$$

and the spectral function

$$A_a(\vec{k}, \sigma, \omega) = -2\,\text{Im}\, G_{a,ret}(\vec{k}, \sigma, \omega) \quad (18)$$

To express our ideas, firstly, we take $H_{d-d} = 0$ in Eq.(4). This Hamiltonian can be diagonalized, or we get

$$G_c(\vec{k}, \sigma, j\omega_n) = \frac{1}{j\omega_n - E_k - |M_{\vec{k}}|^2 / (j\omega_n - \varepsilon_k)} \quad (19)$$

Let $\omega - E_k - |M_{\vec{k}}|^2 / (\omega - \varepsilon_k) = 0$, we find two solutions

$$\omega = \varepsilon_k^{(0,\pm)} = \frac{1}{2}[E_k + \varepsilon_k \pm \sqrt{(E_k - \varepsilon_k)^2 + 4|M_{\vec{k}}|^2}] \quad (20)$$

thus the diagonalized Hamiltonian is

$$H_0 = \sum_{\vec{k}\sigma} \varepsilon_k^{(0,+)} \alpha_{\vec{k}\sigma}^+ \alpha_{\vec{k}\sigma} + \sum_{\vec{k}\sigma} \varepsilon_k^{(0,-)} \beta_{\vec{k}\sigma}^+ \beta_{\vec{k}\sigma} \quad (21)$$

It seems that $a_{\vec{k}\sigma}$ should be taken as either $\alpha_{\vec{k}\sigma}$ or $\beta_{\vec{k}\sigma}$. In fact, $a_{\vec{k}\sigma}$ should be $\beta_{\vec{k}\sigma}$ due to the restriction of the Fermi function $n_F^a(E_B^{\vec{k}})$ in Eq.(15) when we consider the electronic properties of materials, and then $A_a(\vec{k}, \sigma, E_B^{\vec{k}}) = 2\pi\delta(E_B^{\vec{k}} - \varepsilon_k^{(0,-)})$ in this example. Because $\varepsilon_k^{(0,-)} \neq \varepsilon_k$, due to the effect of the overlap matrix element $M_{\vec{k}}$, the electronic structure inside the solid could not be determined directly by ARPES experiments. However, one-band model for $H_0$ could be taken for good metals, in this case, $A_a(\vec{k},\sigma,\omega)$ is just the spectral function of the electron systems in these metals.

For actual materials, especially for some strongly correlated materials, we should consider other interactions $H_{d-d} \neq 0$. In this case, generally speaking, one finds

$$G_c(\vec{k},\sigma, j\omega_n) = \frac{1}{j\omega_n - E_k - |M_{\vec{k}}|^2/(j\omega_n - \varepsilon_k) - \Sigma(\vec{k}, j\omega_n)} \qquad (22)$$

If we denote by $\omega$ the solution of the equation

$$\omega - E_k - |M_{\vec{k}}|^2/(\omega - \varepsilon_k) - \mathrm{Re}\Sigma(\vec{k},\omega) = 0 \qquad (23)$$

there should be two real solutions for $\omega$, $\varepsilon_k^{(+)}$ and $\varepsilon_k^{(-)}$, and the spectral function in regions where $\mathrm{Im}\Sigma(\vec{k},\omega)=0$ can be written as

$$A_c = A_c^{(+)} + A_c^{(-)}$$
$$A_c^{(+)} = \gamma_1(\vec{k},\sigma,\omega) 2\pi\delta(\omega - \varepsilon_k^{(+)}) \qquad (24)$$
$$A_c^{(-)} = \gamma_2(\vec{k},\sigma,\omega) 2\pi\delta(\omega - \varepsilon_k^{(-)})$$

The spectral function in the energy region where $\mathrm{Im}\Sigma(\vec{k},\omega) \neq 0$ can also be observed in the experiments, but it is not our focus in this work since it does not correspond to the quasi-particle spectral function. The spectral function $A_a(\vec{k},\sigma,\omega)$ should be taken as $A_c^{(-)}$. Moreover, $A_c^{(-)}$ is not the exact spectral function inside the solid, and $\varepsilon_k^{(-)}$ is also affected by $M_{\vec{k}}$. This shows that the electronic structures in strongly correlated materials must be determined on the basis of combining ARPES experiment with the solution of the theoretic model. Some models similar to Eq.(4) have been solved as reported in the literatures, such as in these literatures [7-10]. Could they be applied to explain the electronic structures of strongly correlated materials? This has to be discriminated by experiments with Eq.(15).

**4. Summary**

This work derives the formula of photoemission intensity on the two-band model and discusses how to determine the electronic structures in strongly correlated materials by ARPES experiments. Eqs.(1)-(15) do not depend on the following discussion in this work, and these equations show that the effects of the overlap matrix element $M_{\vec{k}}$ must be eliminated when we intend to understand the results of ARPES experiments.

**ACKNOWLEDGMENTS**

The author thanks Nanjing University of Information Science & Technology for financial support.